\documentclass[sigplan,screen]{acmart}
\AtBeginDocument{%
  \providecommand\BibTeX{{%
    \normalfont B\kern-0.5em{\scshape i\kern-0.25em b}\kern-0.8em\TeX}}}

\setcopyright{rightsretained}
\copyrightyear{2022}
\acmYear{2022}
\acmDOI{}

\acmConference[CHI EA '22, Proceedings of the 1st Workshop on Novel Challenges of Safety, Security and Privacy in Extended Reality]{CHI Conference on Human Factors in Computing Systems}{April 29-May 5, 2022}{New Orleans, LA, USA}
\acmBooktitle{CHI Conference on Human Factors in Computing Systems (CHI '22), April 29-May 5, 2022, New Orleans, LA, USA}

\acmPrice{15.00}
\acmDOI{}
\acmISBN{}

\begin{document}

\newcommand{\qingxiao}[1]{{\small\textcolor{blue}{\bf [*** Tasks: #1]}}}
\newcommand{\yun}[1]{{\small\textcolor{orange}{\bf [Yun: #1]}}}

\title{Facing the Illusion and Reality of Safety in Social VR}


\author{Qingxiao Zheng}
\affiliation{%
  \country{School of Information Sciences\\
  University of Illinois at Urbana-Champaign, USA}}
\email{qzheng14@illinois.edu}

\author{Tue Ngoc Do}
\affiliation{%
  \country{Department of Computer Science\\
  University of Illinois at Urbana-Champaign, USA}}
 \email{tuedo2@illinois.edu}

\author{Lingqing Wang}
\affiliation{%
  \country{Department of Industrial Engineering\\Tsinghua University, China}}
 \email{wanglq19@mails.tsinghua.edu.cn}

\author{Yun Huang}
\affiliation{%
 \country{School of Information Sciences\\
  University of Illinois at Urbana-Champaign, USA}}
 \email{yunhuang@illinois.edu}

\renewcommand{\shortauthors}{Zheng, et al.}

\begin{abstract}
The ethical design of social Virtual Reality (VR) is not a new topic, but ``safety'' concerns of using social VR are escalated to a different level given the heat of the Metaverse. For example, it was reported that nearly half of the female-identifying VR participants have had at least one instance of virtual sexual harassment. Feeling safe is a basic human right — in any place, regardless in real or virtual spaces. In this paper, we are seeking to understand the discrepancy between user concerns and designs in protecting users' safety in social VR applications. We study safety concerns on social VR experience first by analyzing Twitter posts, and then synthesize practices on safety protection adopted by four mainstream social VR platforms. We argue that future research and platforms should explore the design of social VR with boundary-awareness.
 
\end{abstract}




\addtolength{\topmargin}{-5.67603pt}

\maketitle

\section{Introduction}
The 3D virtual world, or metaverse~\cite{dionisio20133d}, is broadly defined to be the recent flush of technologies including and related to Virtual Reality (VR) and Augmented Reality (AR), consisting of permanent virtual worlds that persist when we log off. This new dimension of user interaction is likely to spark new social norms and cultural behaviors.

However, attention has been drawn towards the safety, or lack of, features currently existing in commercial VR spaces. For example, a recent news report  ~\cite{Barrett2022} asserted that virtual worlds lack adequate safety precautions; reporting on a group of researchers who recorded a social transgression (e.g. raci-\\sm or sexual harassment) on average once per seven minutes, and often with minors present. Heavy criticism has been raised on some VR front runners, for focusing on growth at the expense of safety and privacy~\cite{dick2021balancing, Lorne2021, blackwell2019harassment}.

Furthermore, considerations of the immersive facet of virtual reality have led some to explore new avenues of online attacks. In comparison to traditional social medias, in which a user has the physical barrier of a screen held at arms length, psychological attacks within these spaces are potentially more traumatic~\cite{Barrett2022, Tanya2021, Sheera2021}. There seems to be less boundaries in VR that can rule and determine what are reasonable, psychologically safe and permissible ways for other people to behave around self and how self will respond when someone steps outside those limits. To this end, we are trying to explore the following questions:

\textbf{RQ1}: \textit{What safety concerns are emerging in social VR platforms?} 

\textbf{RQ2}: \textit{What mechanisms do mainstream social VR products use to protect user safety?}


\section{Related Social VR Research}
Social VR has been defined as “a growing set of multi-user applications that enable people to interact with one another in virtual space through VR headmounted displays”~\cite{kolesnichenko2019understanding}. Today's social VR applications are similar to “collaborative virtual
environments” studied in the late 1990s and early 2000s, which applied virtual reality to support multiple user interactions~\cite{becker1998social}.

There are a set of works mapping social VR landscape and design research by proposing a taxonomy of social VR application design ecology~\cite{zytko2022taxonomy, mcveigh2018s}. Some work specially investigated codes of ethics in the virtual worlds~\cite{adams2018ethics, gray2019ethical, madary2016real}. Some studied social VR from human factor perspective. For example, one study investigated how violations of interpersonal space can cause discomfort in real-world situations~\cite{wilcox2006personal}, and another work learned how users shape pro-social behavior in VR~\cite{mcveigh2019shaping}. 

New questions have arisen from the embodied and spatialized experiences that immersive presence in VR affords in recent years. Harassment in social VR, for instance, has proven particularly fraught, with the effects of abuse magnified by embodied presence~\cite{mcveigh2018s}. However, to the best of our knowledge, research related to safety mechanisms applied in social VR is still very limited.

\section{Method}
We conducted an exploratory study. The seed findings will be expanded with a more comprehensive investigation in the future.

\subsection{Scraping Social Media Posts}
To answer RQ1, we used the Tweepy and PRAW Python packages to scrape recent three months' content from Twitter and Reddit, respectively. These social media platforms were chosen due to their more public natures  and author familiarity. We used search queries of social VR synonyms (e.g., social VR, social virtual reality, virtual reality, virtual worlds) and names of commercial social VR applications~\cite{zytko2022taxonomy}. We ended up with 18,274 data entries and proceeded by manually reviewing all tweets and comments after filtering by keywords. Using a thematic analysis approach~\cite{terry2017thematic}, we synthesized major concerns and discussion topics of social media posts related to social VR.

\subsection{Investigating Mainstream Social VR Platforms}
To answer RQ2, we investigated 14 existing social VR commercial applications previously summarized~\cite{zytko2022taxonomy, jonas2019towards}. Based on number of downloads in across VR app stores of Oculus Store\footnote{https://www.oculus.com/experiences/quest}, Steam VR\footnote{https://store.steampowered.com/vr}, and Sidequest \footnote{https://sidequestvr.com/all-apps}, we chose four social VR apps. They are Rec Room\footnote{https://recroom.com}, which allows users to manipulate a variety of objects and engage with mini-games with other users; VRChat\footnote{https://hello.vrchat.com/}, which is known for non-normative social interactions and performative memes; Horizon Worlds\footnote{https://www.oculus.com/horizon-worlds}, which allows users to explore virtual worlds and engage in content consumption and creation; and AltspaceVR\footnote{https://altvr.com}, which is used for live, virtual events, empowering artists, brands, and businesses. One researcher experienced these social VR apps using an Oculus Quest 2 headset and triggered the safety mechanisms built within to examine these apps' safety protection protocols. Then, one co-author summarized the observations of the safety mechanisms applied by the apps. Additionally, co-authors analyzed the relationship between safety-related design features and the kinds of social interactions they may support or constrain, by reviewing their experiences of onboarding, exploration, and play.

\section{Preliminary Takeaways}
\subsection{Safety Concerns about Social VR (RQ1)}

\subsubsection{Safety of Minors}

A key concern mentioned by multiple users across platforms were the existence of \textbf{minors} in these VR environments. Although users seemed to be split on whether to blame parents or developers for negligence, users unanimously expressed that minors did not belong on the platform, as their safety could not be guaranteed. The underlying reasoning was the perceived accessibility of dangerous/unfitting material. For example, one Twitter user made an analogy to the physical world "\textit{There's a benefit to the physical distance between the average house and the Vegas strip. Harder for your kids to just wander over there.}"

Virtual reality has attracted many young users with chat rooms and popular games, such as Roblox. But whether it can adequately protect children is a major concern. Many users of Horizon Worlds have reported that Facebook, now Meta, does not sufficiently enforce the age restriction, which leads to worries for minors' safety and a negative adult user experience. These problems have been widely discussed and addressed on social media. However, some platforms in virtual reality spaces still lag behind common concerns, lacking pre-existing methods such as parental controls and guard-\\rails for young users, to the frustrations of some community members.

\subsubsection{Distrust of Corporations}

Another concern is distrust. Specifically, regarding the track record of Meta, there seemed to be some pushback with Reddit users feeling uncomfortable plugging in at all. One Reddit user remarked "\textit{I can assure you I will not be using it} [Oculus]\textit{...Zuckerberg has revealed over and over he doesn't [care (replaced word for polite language)] about safety or privacy.}"

There also seems to be a disdain for Meta's current moderation systems. Regarding \textbf{sexual harassment }and \textbf{abusive behavior}, one commentor expresses in regards to a blocking feature (at the time unimplemented), "\textit{I would not be surprised if Facebook empowers users even in this minor way. You know they aren't gonna bother banning creeps.}"

In December 2021, a beta tester of Horizon Worlds wrote about being virtually groped by another user, creating quite a stir. After that, while VR platforms are claiming to develop different tools and improve their accessibility to protect users, more harassment cases are reported. Sexual harassment exists in all kinds of digital spaces. However, the immersive, multi-sensory and ‘realistic’ nature of VR triggers stronger emotional responses, thus even worse repercussions for users in such circumstances. 

\subsubsection{Humiliating Behaviors.}

Currently, the public seems to largely rehash many known concerns with other social media without considering the additional dimension and complications a virtual reality can bring. For example, regarding a now-famous incident of a woman being groped while in VR, some commentors were dismissive of her experiences, suggesting that woman just "turn it off", and falsely equalizing the experience with milder internet norms such as teabagging (the act of repeatedly crouching). As more incidents were logged with these technologies and social space, and new norms and cultures are yet established, it remains to be seen where public opinions start to lean in terms of user safety in the virtual worlds.

\subsubsection{Integration with the \textit{Reality}}

Many users have been discussing that Social VR platforms might blur the boundary between reality and the virtual world. Despite users’ worries that the VR world may lead to addiction easily, causing more serious \textbf{isolation} from real-world and cognitive problems. It also raises concerns when integrating into the real world. For example, there are several news reports on the physical injury in the real-world caused by body movement in the virtual world, when users immerse themselves in the playing. 

Some other emerging issues have also caught public attention. One is the \textbf{sleeping phenomenon}. Widely discussed on YouTube channels, there is a unique phenomenon that users regularly sleep on virtual platforms despite all the inconveniences in real world. For example, the headset might hinder head movement while sleeping, and the cable might tie around users' neck. Reasons for sleeping in a social VR platform vary from social satisfaction when sleeping with friends to avoiding the real-world noisy environment. Though users make the trade-off to do so, they are generally not clear about the potential consequence.

Another emerging issue is phantom touch, or more broadly speaking, \textbf{phantom sense}, which means users feel the sensation as if they are their avatars, without any haptic technology. One Reddit user wrote "\textit{when someone headpats me (well, my avatar, but to me, they're the same thing... ) it gives me a tingly feeling on my head ...}" There are many explanations and guides to help users develop phantom touch in Reddit and other discussion forums. Some users claim they naturally gain phantom sense, and other provides various methods to trick the mind into that. For example, one user in the Steam community recommended asking a friend to pet users' face in VR, suggesting that "\textit{area most sensitive for most people is the small part between the upper lips and the bottom of the nose - especially to the left and right of the center.}"  While this phenomenon is not clear, it adds exciting layers to user experience in virtual world. Many users are trying to practice and keep the feelings of phantom touches in VR. However, many users also express their worries about whether they can still differentiate between reality and the virtual environment. After all, having a bad experience for the avatars, like being shoved or stabbed, might not be fun.

In summary, we found that safety concerns on social media include "intensified old concerns in the new world." Harassment and abusive behaviors, distrust of user privacy, content appropriateness for children etc., have long been discussed in previous social platforms. And now, they exacerbate users' concerns because of the double-edged, more authentic experience that social VR platforms promise to provide, which would also bring more real lousy experience at the same time. 

Also, some "new concerns in the new world" have recently become salient, such as problems caused by the separation of body and mind in real and virtual worlds, and emerging issues like sleeping phenomenon and phantom sense. How work and lifestyles would be reshaped, and what that might impact user  psychological and physical safety are not yet addressed extensively.

\subsection{Designs Features for User Safety (RQ2)}
In a prior work, we advocate raising "boundary-awareness" when designing AI-based interactive systems to support hu-\\man-human interaction~\cite{zheng2022ux}. When reviewing safety mechanisms adopted by main stream social VR platforms, we identified six major social boundary settings, four safety protection mechanisms, and community protocols.

\subsubsection{Setting Social Boundaries}
We found six boundary related design practices to keep users comfortable with social distance when hanging out in social VR.

\textbf{\textit{(1) Intimacy Proxemic}.} Hall's personal space theory describes that there are four zones in personal space~\cite{hall1968proxemics}. Recently, Horizon Worlds released the personal boundary feature with a default setting making users feel like an almost 4-foot distance between their avatar and others. Similarly, Rec Room has an adjustable "personal space bubble," an invisible space that allows users to control how close other players interact with their avatars from close, medium, to large. We also found a similar function in AltspaceVR.

\textbf{\textit{(2) Intimacy Rank}}. When playing VRchat, users' trust feeds into a "trust rank" with several levels, i.e., trusted user, known user, user, new user, visitor, and nuisance.

\textbf{\textbf{}\textit{(3) Social Spaces}}. Rec Room has a dorm room, user-created rooms, and clubs. The dorm room is a completely private space. Room owners of user-created rooms and clubs are fully responsible for ensuring their rooms do not devolve into a toxic mess, even if they are going on vacations in the real world. We found that local expectations are set for conduct in public spaces. For example, it must be private if a club involves sexual themes.

\textbf{\textit{(4) Image Control}}. Users can adjust both their own avatars' identities, such as gender, race, and voice, and select to hide or show specific users' avatars. For example, in VRchat, when a user encounters a user that, despite their higher Trust Rank, is wearing an avatar they feel uncomfortable, they can choose "hide avatar" in their social panel.

\textbf{\textit{(5) Trust System}}. Users can adjust control settings to maintain desirable social boundaries with others. For example, VRChat trust and safety system can keep users safe from nuisance by adjusting microphones, screen-space shaders, loud sounds, or visually noisy or malicious particle effects.

\textbf{\textit{(6) User Demographics}}. Age limits are set in social VR. For example, Rec Room enables a junior account managed by an account owned by the parent or guardian for players under 13 years of age, which come with safety features to prevent children from disclosing their identifiable information in the virtual world. It also allows age-based matchmaking to recommend players close to their age range.

\subsubsection{Quick Reactions When Boundaries are Crossed}
Different safety self-protection mechanisms were designed to allow users to react accordingly when their social boundaries are violated in the virtual world, ranging from reactions and safe zones to safety reports. 

\textbf{\textit{Safety Reactions.}} 
Simple communication gestures and shortcuts allowed quick-action remediation in challenging situations. For example, Rec Room's 'talk to the hand' and pointing gestures can instantly trigger users' comfort and moderation menu, allowing users to mute, block, vote kick, and report the player. These simple mechanisms enable users to report a hurting experience instantly without interrupting or degrading their experience.

\textbf{\textit{Safe Mode.}} VRchat has a shortcut called "safe mode," which can immediately disable all features on all users around. Horizon worlds offer a one-touch button that can quickly remove users from a situation. Similarly, AltspaceVR has a "title screen" feature, which immediately drops users back into the command center, removing their avatars from the previous space and scene. 

\textbf{\textit{Safety Reports.}} Users can report users, rooms or event places that violate codes of conduct or ask for extra support.

\subsubsection{Consent and Codes of Conduct}
All applications have informed Consent before using entering the application. The informed consent contents regarding health and safety warnings range from the immersive effects on users' psychological status, a notice of risky content during social interactions, and user privacy. Some applications, for example, Rec Room, also require users to finish a tutorial of codes of conduct in the virtual environment; these codes were introduced to newcomers and repeated displayed to members of the communities.

\section{Opportunities for Future Research}
Initial results reveal many old safety concerns in the new virtual world (e.g., user trust, abusive language and behavior, threats to minors) which current design features have addressed. However, some new issues, such as the "phantom sense," are not considered by any reviewed social VR apps. There are many novel safety facets of this emerging design space of social VR. Below, we propose some topics to explore.

\subsection{Re-defining User Safety in Social VR}
Do we need to redefine safety in social VR? VR has many shared characteristics with our experiences in real life. Because players can disguise themselves into another human, animal, or object and use transportation behaviors, they may have different social boundaries and appropriateness definitions than as usual. Due to its nascent stage and creative formats, it is yet to investigate the social norms being applied in the new world. One approach is distinguishing universal rules and consequences for real-world and virtual-world violations and harassment.

\subsection{Boundary-Awareness for Safter Social Systems}
Boundary-aware design is an emerging and under-explored research area\cite{zheng2022ux}. Our findings on social VR safety mechanisms demonstrate the dynamics of boundary considerations in social systems. Studying how these systems consider various social boundaries and the interactions occurring at different levels of boundaries, may add value to the human-human interactions through these social technologies. For example, building granular controls on different boundary levels (e.g., on information sharing, emotional bou-\\ndaries, differentiating reality and virtual, persona building, to name a few) may help foster safety in social VR spaces. Researchers can also develop a boundary-aware framework to support the design process and evaluations with such awareness.

\subsection{Values of Platforms, Creators, and Users}
Building safe and responsible VR ecology will have a competitive edge in the virtual world.  For platforms, toolkits providers that have the autonomy to limit content creators' production capabilities may easily restrict creators' development capabilities by restraining risky developer support. Creators, such as those in the Decentraland, a first-ever platform to create, explore, and trade land in the virtual world owned by its users, have high autonomy in the "fully decentralized world." Platforms can hardly control the content created. For users, we found community consensuses and discussions regarding safety in social VR are not fully developed, perhaps due to its novelty. One possible approach perhaps can be community-based safety reports, where bystanders can intervene and stop a violent attack by one person on another~\cite{slater2013bystander}. A shared understanding is needed and we posit that more future studies could uncover more fruitful understandings.



\bibliographystyle{ACM-Reference-Format}
\bibliography{references}


\begin{thebibliography}{21}


\ifx \showCODEN    \undefined \def \showCODEN     #1{\unskip}     \fi
\ifx \showDOI      \undefined \def \showDOI       #1{#1}\fi
\ifx \showISBNx    \undefined \def \showISBNx     #1{\unskip}     \fi
\ifx \showISBNxiii \undefined \def \showISBNxiii  #1{\unskip}     \fi
\ifx \showISSN     \undefined \def \showISSN      #1{\unskip}     \fi
\ifx \showLCCN     \undefined \def \showLCCN      #1{\unskip}     \fi
\ifx \shownote     \undefined \def \shownote      #1{#1}          \fi
\ifx \showarticletitle \undefined \def \showarticletitle #1{#1}   \fi
\ifx \showURL      \undefined \def \showURL       {\relax}        \fi
\providecommand\bibfield[2]{#2}
\providecommand\bibinfo[2]{#2}
\providecommand\natexlab[1]{#1}
\providecommand\showeprint[2][]{arXiv:#2}

\bibitem[Adams et~al\mbox{.}(2018)]%
        {adams2018ethics}
\bibfield{author}{\bibinfo{person}{Devon Adams}, \bibinfo{person}{Alseny Bah},
  \bibinfo{person}{Catherine Barwulor}, \bibinfo{person}{Nureli Musaby},
  \bibinfo{person}{Kadeem Pitkin}, {and} \bibinfo{person}{Elissa~M Redmiles}.}
  \bibinfo{year}{2018}\natexlab{}.
\newblock \showarticletitle{Ethics emerging: the story of privacy and security
  perceptions in virtual reality}. In \bibinfo{booktitle}{\emph{Fourteenth
  Symposium on Usable Privacy and Security (SOUPS 2018)}}.
  \bibinfo{pages}{427--442}.
\newblock


\bibitem[Barrett and Douglas(2019)]%
        {Barrett2022}
\bibfield{author}{\bibinfo{person}{Maura Barrett} {and} \bibinfo{person}{Forte
  Douglas}.} \bibinfo{year}{2019}\natexlab{}.
\newblock \bibinfo{booktitle}{\emph{Metaverse virtual worlds lack adequate
  safety precautions}}.
\newblock


\bibitem[Basuarchive(2021)]%
        {Tanya2021}
\bibfield{author}{\bibinfo{person}{Tanya Basuarchive}.}
  \bibinfo{year}{2021}\natexlab{}.
\newblock \bibinfo{booktitle}{\emph{The metaverse has a groping problem
  already}}.
\newblock


\bibitem[Becker and Mark(1998)]%
        {becker1998social}
\bibfield{author}{\bibinfo{person}{Barbara Becker} {and}
  \bibinfo{person}{Gloria Mark}.} \bibinfo{year}{1998}\natexlab{}.
\newblock \showarticletitle{Social conventions in collaborative virtual
  environments}. In \bibinfo{booktitle}{\emph{Proceedings of Collaborative
  Virtual Environments}}. \bibinfo{pages}{17--19}.
\newblock


\bibitem[Blackwell et~al\mbox{.}(2019)]%
        {blackwell2019harassment}
\bibfield{author}{\bibinfo{person}{Lindsay Blackwell}, \bibinfo{person}{Nicole
  Ellison}, \bibinfo{person}{Natasha Elliott-Deflo}, {and} \bibinfo{person}{Raz
  Schwartz}.} \bibinfo{year}{2019}\natexlab{}.
\newblock \showarticletitle{Harassment in social VR: Implications for design}.
  In \bibinfo{booktitle}{\emph{2019 IEEE Conference on Virtual Reality and 3D
  User Interfaces (VR)}}. IEEE, \bibinfo{pages}{854--855}.
\newblock


\bibitem[Dick(2021)]%
        {dick2021balancing}
\bibfield{author}{\bibinfo{person}{Ellysse Dick}.}
  \bibinfo{year}{2021}\natexlab{}.
\newblock \bibinfo{booktitle}{\emph{Balancing User Privacy and Innovation in
  Augmented and Virtual Reality}}.
\newblock \bibinfo{type}{{T}echnical {R}eport}.
  \bibinfo{institution}{Information Technology and Innovation Foundation}.
\newblock


\bibitem[Dionisio et~al\mbox{.}(2013)]%
        {dionisio20133d}
\bibfield{author}{\bibinfo{person}{John David~N Dionisio},
  \bibinfo{person}{William G~Burns III}, {and} \bibinfo{person}{Richard
  Gilbert}.} \bibinfo{year}{2013}\natexlab{}.
\newblock \showarticletitle{3D virtual worlds and the metaverse: Current status
  and future possibilities}.
\newblock \bibinfo{journal}{\emph{ACM Computing Surveys (CSUR)}}
  \bibinfo{volume}{45}, \bibinfo{number}{3} (\bibinfo{year}{2013}),
  \bibinfo{pages}{1--38}.
\newblock


\bibitem[Gray and Chivukula(2019)]%
        {gray2019ethical}
\bibfield{author}{\bibinfo{person}{Colin~M Gray} {and}
  \bibinfo{person}{Shruthi~Sai Chivukula}.} \bibinfo{year}{2019}\natexlab{}.
\newblock \showarticletitle{Ethical mediation in UX practice}. In
  \bibinfo{booktitle}{\emph{Proceedings of the 2019 CHI Conference on Human
  Factors in Computing Systems}}. \bibinfo{pages}{1--11}.
\newblock


\bibitem[Hall et~al\mbox{.}(1968)]%
        {hall1968proxemics}
\bibfield{author}{\bibinfo{person}{Edward~T Hall}, \bibinfo{person}{Ray~L
  Birdwhistell}, \bibinfo{person}{Bernhard Bock}, \bibinfo{person}{Paul
  Bohannan}, \bibinfo{person}{A~Richard Diebold~Jr}, \bibinfo{person}{Marshall
  Durbin}, \bibinfo{person}{Munro~S Edmonson}, \bibinfo{person}{JL Fischer},
  \bibinfo{person}{Dell Hymes}, \bibinfo{person}{Solon~T Kimball},
  {et~al\mbox{.}}} \bibinfo{year}{1968}\natexlab{}.
\newblock \showarticletitle{Proxemics [and comments and replies]}.
\newblock \bibinfo{journal}{\emph{Current anthropology}} \bibinfo{volume}{9},
  \bibinfo{number}{2/3} (\bibinfo{year}{1968}), \bibinfo{pages}{83--108}.
\newblock


\bibitem[Jonas et~al\mbox{.}(2019)]%
        {jonas2019towards}
\bibfield{author}{\bibinfo{person}{Marcel Jonas}, \bibinfo{person}{Steven
  Said}, \bibinfo{person}{Daniel Yu}, \bibinfo{person}{Chris Aiello},
  \bibinfo{person}{Nicholas Furlo}, {and} \bibinfo{person}{Douglas Zytko}.}
  \bibinfo{year}{2019}\natexlab{}.
\newblock \showarticletitle{Towards a taxonomy of social vr application
  design}. In \bibinfo{booktitle}{\emph{Extended Abstracts of the Annual
  Symposium on Computer-Human Interaction in Play Companion Extended
  Abstracts}}. \bibinfo{pages}{437--444}.
\newblock


\bibitem[Kolesnichenko et~al\mbox{.}(2019)]%
        {kolesnichenko2019understanding}
\bibfield{author}{\bibinfo{person}{Anya Kolesnichenko}, \bibinfo{person}{Joshua
  McVeigh-Schultz}, {and} \bibinfo{person}{Katherine Isbister}.}
  \bibinfo{year}{2019}\natexlab{}.
\newblock \showarticletitle{Understanding emerging design practices for avatar
  systems in the commercial social vr ecology}. In
  \bibinfo{booktitle}{\emph{Proceedings of the 2019 on Designing Interactive
  Systems Conference}}. \bibinfo{pages}{241--252}.
\newblock


\bibitem[Lorne(2021)]%
        {Lorne2021}
\bibfield{author}{\bibinfo{person}{Fade Lorne}.}
  \bibinfo{year}{2021}\natexlab{}.
\newblock \bibinfo{booktitle}{\emph{What Is Virtual Reality, And How Can It Be
  Used In The Workplace?}}
\newblock


\bibitem[Madary and Metzinger(2016)]%
        {madary2016real}
\bibfield{author}{\bibinfo{person}{Michael Madary} {and}
  \bibinfo{person}{Thomas~K Metzinger}.} \bibinfo{year}{2016}\natexlab{}.
\newblock \showarticletitle{Real virtuality: a code of ethical conduct.
  Recommendations for good scientific practice and the consumers of
  VR-technology}.
\newblock \bibinfo{journal}{\emph{Frontiers in Robotics and AI}}
  \bibinfo{volume}{3} (\bibinfo{year}{2016}), \bibinfo{pages}{3}.
\newblock


\bibitem[McVeigh-Schultz et~al\mbox{.}(2019)]%
        {mcveigh2019shaping}
\bibfield{author}{\bibinfo{person}{Joshua McVeigh-Schultz},
  \bibinfo{person}{Anya Kolesnichenko}, {and} \bibinfo{person}{Katherine
  Isbister}.} \bibinfo{year}{2019}\natexlab{}.
\newblock \showarticletitle{Shaping pro-social interaction in VR: an emerging
  design framework}. In \bibinfo{booktitle}{\emph{Proceedings of the 2019 CHI
  Conference on Human Factors in Computing Systems}}. \bibinfo{pages}{1--12}.
\newblock


\bibitem[McVeigh-Schultz et~al\mbox{.}(2018)]%
        {mcveigh2018s}
\bibfield{author}{\bibinfo{person}{Joshua McVeigh-Schultz},
  \bibinfo{person}{Elena M{\'a}rquez~Segura}, \bibinfo{person}{Nick Merrill},
  {and} \bibinfo{person}{Katherine Isbister}.} \bibinfo{year}{2018}\natexlab{}.
\newblock \showarticletitle{What's It Mean to" Be Social" in VR? Mapping the
  Social VR Design Ecology}. In \bibinfo{booktitle}{\emph{Proceedings of the
  2018 ACM Conference Companion Publication on Designing Interactive Systems}}.
  \bibinfo{pages}{289--294}.
\newblock


\bibitem[Sheera and Kellen(2021)]%
        {Sheera2021}
\bibfield{author}{\bibinfo{person}{Frenkel Sheera} {and}
  \bibinfo{person}{Browning Kellen}.} \bibinfo{year}{2021}\natexlab{}.
\newblock \bibinfo{booktitle}{\emph{The Metaverse’s Dark Side: Here Come
  Harassment and Assaults}}.
\newblock


\bibitem[Slater et~al\mbox{.}(2013)]%
        {slater2013bystander}
\bibfield{author}{\bibinfo{person}{Mel Slater}, \bibinfo{person}{Aitor Rovira},
  \bibinfo{person}{Richard Southern}, \bibinfo{person}{David Swapp},
  \bibinfo{person}{Jian~J Zhang}, \bibinfo{person}{Claire Campbell}, {and}
  \bibinfo{person}{Mark Levine}.} \bibinfo{year}{2013}\natexlab{}.
\newblock \showarticletitle{Bystander responses to a violent incident in an
  immersive virtual environment}.
\newblock \bibinfo{journal}{\emph{PloS one}} \bibinfo{volume}{8},
  \bibinfo{number}{1} (\bibinfo{year}{2013}), \bibinfo{pages}{e52766}.
\newblock


\bibitem[Terry et~al\mbox{.}(2017)]%
        {terry2017thematic}
\bibfield{author}{\bibinfo{person}{Gareth Terry}, \bibinfo{person}{Nikki
  Hayfield}, \bibinfo{person}{Victoria Clarke}, {and} \bibinfo{person}{Virginia
  Braun}.} \bibinfo{year}{2017}\natexlab{}.
\newblock \showarticletitle{Thematic analysis}.
\newblock \bibinfo{journal}{\emph{The SAGE handbook of qualitative research in
  psychology}}  \bibinfo{volume}{2} (\bibinfo{year}{2017}),
  \bibinfo{pages}{17--37}.
\newblock


\bibitem[Wilcox et~al\mbox{.}(2006)]%
        {wilcox2006personal}
\bibfield{author}{\bibinfo{person}{Laurie~M Wilcox}, \bibinfo{person}{Robert~S
  Allison}, \bibinfo{person}{Samuel Elfassy}, {and} \bibinfo{person}{Cynthia
  Grelik}.} \bibinfo{year}{2006}\natexlab{}.
\newblock \showarticletitle{Personal space in virtual reality}.
\newblock \bibinfo{journal}{\emph{ACM Transactions on Applied Perception
  (TAP)}} \bibinfo{volume}{3}, \bibinfo{number}{4} (\bibinfo{year}{2006}),
  \bibinfo{pages}{412--428}.
\newblock


\bibitem[Zheng et~al\mbox{.}(2022)]%
        {zheng2022ux}
\bibfield{author}{\bibinfo{person}{Qingxiao Zheng}, \bibinfo{person}{Yiliu
  Tang}, \bibinfo{person}{Yiren Liu}, \bibinfo{person}{Weizi Liu}, {and}
  \bibinfo{person}{Yun Huang}.} \bibinfo{year}{2022}\natexlab{}.
\newblock \showarticletitle{UX Research on Conversational Human-AI Interaction:
  A Literature Review of the ACM Digital Library}.
\newblock \bibinfo{journal}{\emph{arXiv preprint arXiv:2202.09895}}
  (\bibinfo{year}{2022}).
\newblock


\bibitem[Zytko et~al\mbox{.}(2022)]%
        {zytko2022taxonomy}
\bibfield{author}{\bibinfo{person}{Douglas Zytko}, \bibinfo{person}{Ryan
  Handley}, \bibinfo{person}{Bert Guerra}, {and} \bibinfo{person}{Rukkmini
  Goli}.} \bibinfo{year}{2022}\natexlab{}.
\newblock \showarticletitle{A Taxonomy of Social VR Design}.
\newblock \bibinfo{journal}{\emph{arXiv preprint arXiv:2201.02253}}
  (\bibinfo{year}{2022}).
\newblock


\end{thebibliography}

\appendix

\end{document}